# Quantitative Analysis of AI-Generated Texts in Academic Research: A Study of AI Presence in Arxiv Submissions using AI Detection Tool

Arslan Akram[1, 2]

[1] Department of Computer Science, Faculty of Computer Science and Information Technology, The Superior University, Lahore, 54000, Pakistan

[2] MLC Lab, Maharban House, House # 209, Zafar Colony, Okara, 56300, Pakistan

[*]Corresponding Author: Arslan Akram. Email: PHCS-F21-014@gmail.com

**Abstract:** Many people are interested in ChatGPT since it has become a prominent AIGC model that provides high-quality responses in various contexts, such as software development and maintenance. Misuse of ChatGPT might cause significant issues, particularly in public safety and education, despite its immense potential. The majority of researchers choose to publish their work on Arxiv. The effectiveness and originality of future work depend on the ability to detect AI components in such contributions. To address this need, this study will analyze a method that can see purposely manufactured content that academic organizations use to post on Arxiv. For this study, a dataset was created using physics, mathematics, and computer science articles. Using the newly built dataset, the following step is to put originality.ai through its paces. The statistical analysis shows that Originality.ai is very accurate, with a rate of 98%.

**Keywords:** Artificial Intelligence (AI), ChatGPT, Originality.AI, AI detection, arXiv, Academic Integrity, BERT Model, Binary Classification, Natural Language Processing, Machine Learning, Text Generation, Confusion Matrix, AI generated content detection, large language models, Classification, Text evaluation.

## 1. Introduction

Since ChatGPT's release, artificial intelligence has impacted developments in NLU and NG [1]. Academic journals are among the many sectors that have felt the effects of ChatGPT's influence. Academic discourse and approaches increasingly include AI as it advances. An increase in AI development affects the technical research articles on arXiv, and this study examines this trend in detail. Natural language generation (NLG) models developed more recently have significantly improved the control, variety, and quality of text generated by machines. Phishing [1], disinformation [2], fraudulent product reviews [3], academic dishonesty [4], and toxic spam all take advantage of NLG models' ability to generate novel, manipulable, human-like text at breakneck speeds and efficiencies. Generative models like ChatGPT have recently attracted much attention due to their ability to produce material resembling human writing, images, and more. You can train ChatGPT, an OpenAI-developed variation of the widely used GPT-3 language model, to generate conversational text, translate text, and even create new languages [5]. There still needs to be a straightforward way to tell machine-written text from human-written content, despite generative models like ChatGPT have come a long way in producing natural language. This is true even though ChatGPT and other generative models. When it comes to content moderation and other similar applications, this is crucial for detecting and removing harmful information and automated spam [6].

Figure 1 demonstrates how arXiv distributes different papers submitted after 2019 to 2023. Other colors or bars denote categories, while the vertical axis shows the number of documents in each group. This graph details academic research interests and trends over that period. Categories with higher activity may imply popularity or academic commitment. The proportional number of submissions in different sorts of work also suggests new intellectual orientations. Combining this data with prior patterns and external variables may help explain academic research's shifting character and new paths in particular sectors [7].

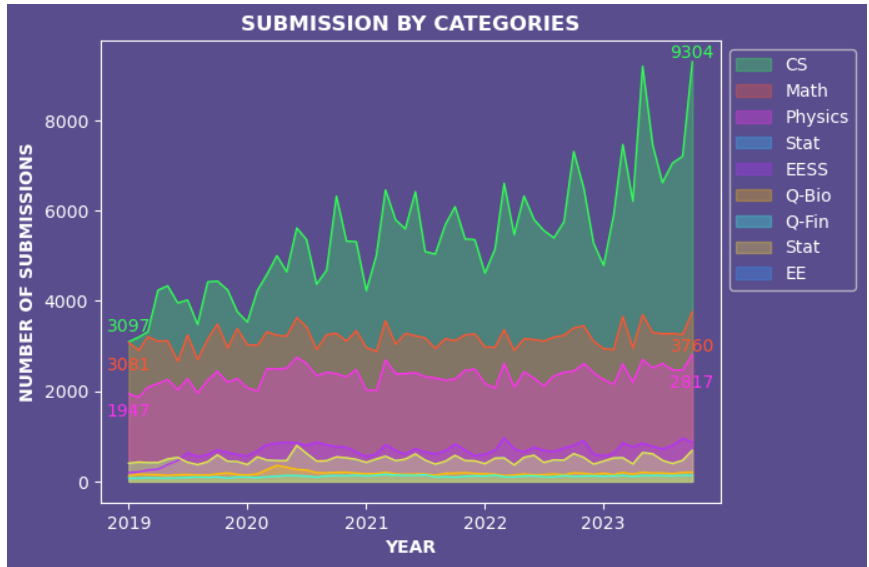

**Figure 1:** Paper Submitted by Categories on Arxiv After the Year 2019. [7]

From 2019 to 2023, the data shows that the top three categories considerably grew in published papers. The largest significant gain was in computer science (200.42%), followed by physics (44.68%) and mathematics (22.04%). There may be a connection between these exponential increases and the use of artificial intelligence writing tools like ChatGPT, which could have allowed for the faster creation of academic publications in specific domains. This theory is in keeping with the general tendency of new technology impacting the way academic journals publish their articles, suggesting that artificial intelligence (AI) may have a revolutionary effect on the quality of research published in these fields.

**Table 1:** Influence of AI text generation tools on arxiv submissions.

| Primary Category | Published Papers in January 2019 | Published Papers in November 2023 | Percentage Increase |
|---|---|---|---|
| Computer Science | 3097 | 9304 | 200.42% |
| Physics | 1947 | 2817 | 44.68% |
| Mathematics | 3081 | 3760 | 22.04% |

Generative models like ChatGPT have recently attracted much attention due to their ability to produce material resembling human writing, images, and more. You can train ChatGPT, an OpenAI-developed variation of the widely used GPT-3 language model, to generate conversational text, translate text, and even create new languages [8]. There still needs to be a straightforward way to tell machine-written text from human-written content, even though generative models like ChatGPT have come a long way in producing language that sounds natural. This is true even though ChatGPT and other generative models. When it comes to content moderation and other similar applications, this is crucial for detecting and removing harmful information and automated spam [9].

The purpose of this study is to quantitatively examine the originality.ai AI-generated text identification tool using a dataset that the researchers have created themselves from arxiv submissions. In order to achieve this, the researchers will scour arxiv.org for literature covering three distinct fields. Unlike previously published studies, this one makes use of a diverse array of text sizes, formats, and organizational patterns. Putting the instrument through its pace and documenting its results is the next step. Presented in the following bullet points are the key points drawn from the study's summary.

- Collecting articles of three different fields from arxiv submissions and form a dataset to use for analysis.
- Reporting performance of originality.ai for detection of AI content in arxiv submissions.

Here is how this article is organized: While Section 2 provides a brief overview of the pertinent literature, Section 4 delves into the results and observations. Finally, the study's conclusions and future research directions are discussed in the part that concludes the work.

## 2. Literature Review

### 2.1 Evolution of AI in Research

The growth of AI in research has been significant. Initially, it was used for basic tasks, but has grown to handle more complex jobs due to advanced algorithms and more computing power [10]. AI can now analyze extremely large sets of data, identify patterns, and even generate new ideas [11]. Important studies in deep learning have shown AI value in more than just data analysis [12], [13]. AI is now used in many research areas [14], [15], speeding up work and leading to new discoveries and innovations.

### 2.2 AI in Scientific Publishing

AI's role in scientific publishing has evolved significantly. It's now a key helper in the writing process, with tools like ChatGPT and Grammarly aiding researchers in writing and editing their work [16]. For instance, AI helps from the first draft of writing to make it more structured. Over the years AI writing has become more accurate and reliable for scientific publications [17].

### 2.3 About the Study on AI in arXiv

This study has looked at how AI influences both the numbers and quality of papers on arXiv. This study includes use of Originality.AI's AI detection tool to check how many arXiv papers are likely to be written by AI. This study helps us understand how AI is changing academic research and shows the need for strong methods to make sure these papers are original and trustworthy.

## 3. Material and Methods

The Originaliry.AI's AI detection model is trained on a million pieces of textual data labeled "human-generated" or "AI-generated", during testing, test the model on documents generated by various artificial intelligence models, including GPT-3, GPT-J, and GPT-NEO (20 thousand data points each). And the result is that the model successfully identified 94.06% of the text created by GPT-3, 94.14% of text written by GPT-J, and 95.64% of text generated by GPT-Neo. The results show that the more powerful the models like GPT-J/3, the harder it is for the model to recognize that the human or AI is writing [18], [19]. After training, the model takes text as input and determines whether it is likely to have been generated by AI or not.

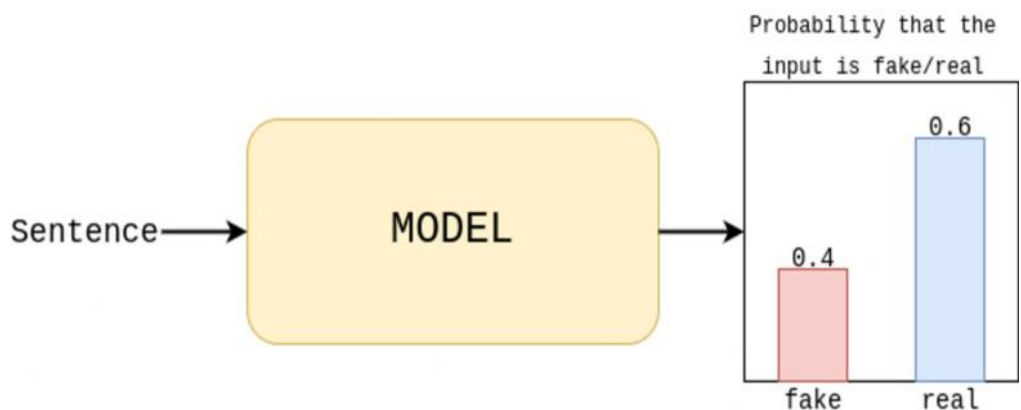

**Figure 2:** Implementation framework for analysis

Originality.AI's AI Detection Tool is an advanced tool that can distinguish AI-generated content from human-written text. It uses a special version of the BERT model [20] that's good at finding AI written content. Originality.AI's AI Detection Tool can accurately identify AI-written text with more than 98% accuracy. This high accuracy comes from thorough testing and updates to keep up with new AI writing models. The tool has been carefully developed and improved to address the increasing use of AI in writing content [21], [22].

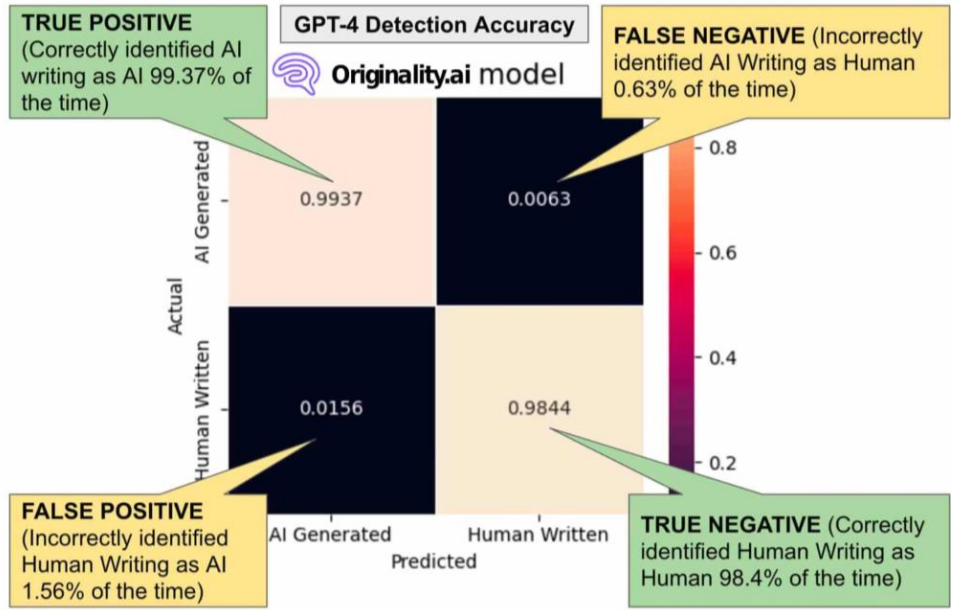

**Figure 3:** Confusion matrix on a gpt-4 human dataset test[21]

The analysis involved using Originality.AI's AI detection tool to distinguish between human and AI-generated content. This tool employs advanced AI algorithms to detect potential AI writing, providing a quantitative measure of AI involvement in these papers. The analysis also included a validation process to assess the tool's accuracy, with a focus on minimizing false positives [21].

*3.1 Dataset Collection*

For this study, we gathered data from 13,000 research papers from the arXiv database. We chose this collection because it covers a wide range of academic topics, which is useful for understanding AI's impact on academic writing. The papers we looked at were chosen based on when they were published, specifically to see how AI has influenced writing since the introduction of ChatGPT.

*3.1  Selection Criteria*

The dataset comprises 60,000 papers initially scraped from the arXiv database, which is a repository for research papers across various fields [23]. The selection criteria involved choosing papers based on relevance to the study's focus on AI's impact on academic writing. A subset of 13,000 papers was then filtered from this larger pool, considering various fields and criteria detailed in our research.

*3.2  Size and Characteristics*

The final dataset consists of 13,000 arXiv papers, offering a diverse and comprehensive collection of academic research. These papers cover multiple disciplines, providing a broad spectrum of research topics and methodologies. The size of the dataset ensures statistical significance for the analysis and enables the exploration of trends across different academic fields.

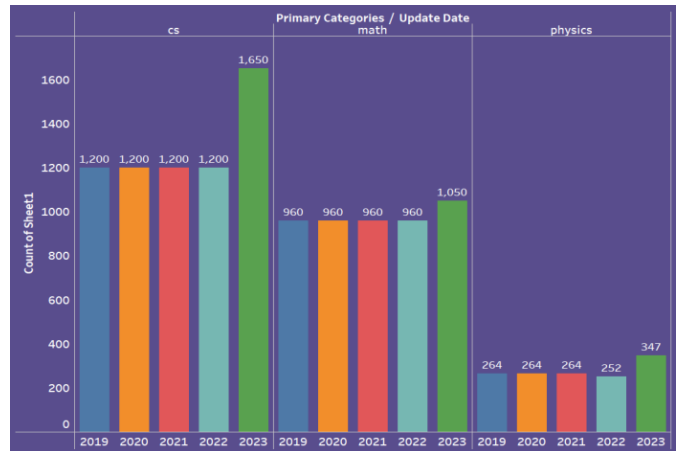

**Figure 4:** Total papers downloaded from arxiv.org

### 3.2 Data Preprocessing
Data preprocessing involved several steps to ensure the quality and relevance of the papers. This included cleaning the data to remove incomplete or irrelevant entries, normalizing the formats for consistency, and categorizing the papers based on specific fields and topics pertinent to the study. Preprocessing was essential to facilitate accurate analysis using the Originality.AI tool.

### 4. Results and Discussions
Using the Originlaiyty.AI's AI detection tool, we scored the papers for AI score and noticed a clear increase in AI-written papers. This method helps us see how AI tools are being used more in academic writing.

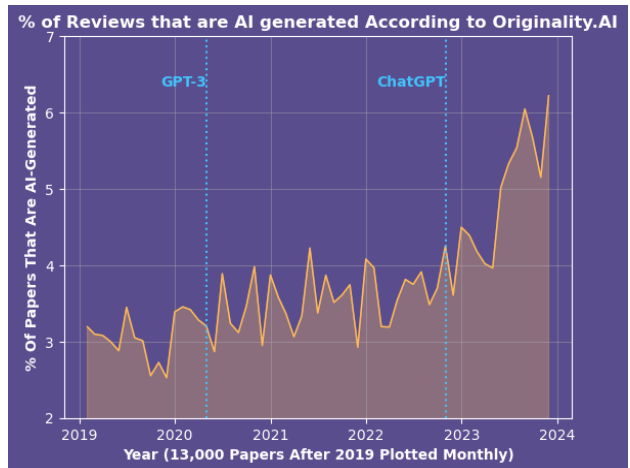

**Figure 5:** Percentage of paper that are likely generated by ai after year 2019 scored by originality's detection tool

The graph shed light on how much AI has been used over time for technical writing. After ChatGPT came out in November 2022, there was a noticeable increase in AI written papers, from 3.61% to 6.22% in a year. This shows that the use of tools like ChatGPT in writing is becoming more common, as seen by the increase in papers with high AI scores after the launch. This suggests that AI tools are affecting the way academic writing is done. This upward trend raised one more question regarding how AI is affecting writing in different categories. To explore the impact of AI across Computer Science, Mathematics, and physics a detailed analysis of AI score has been done.

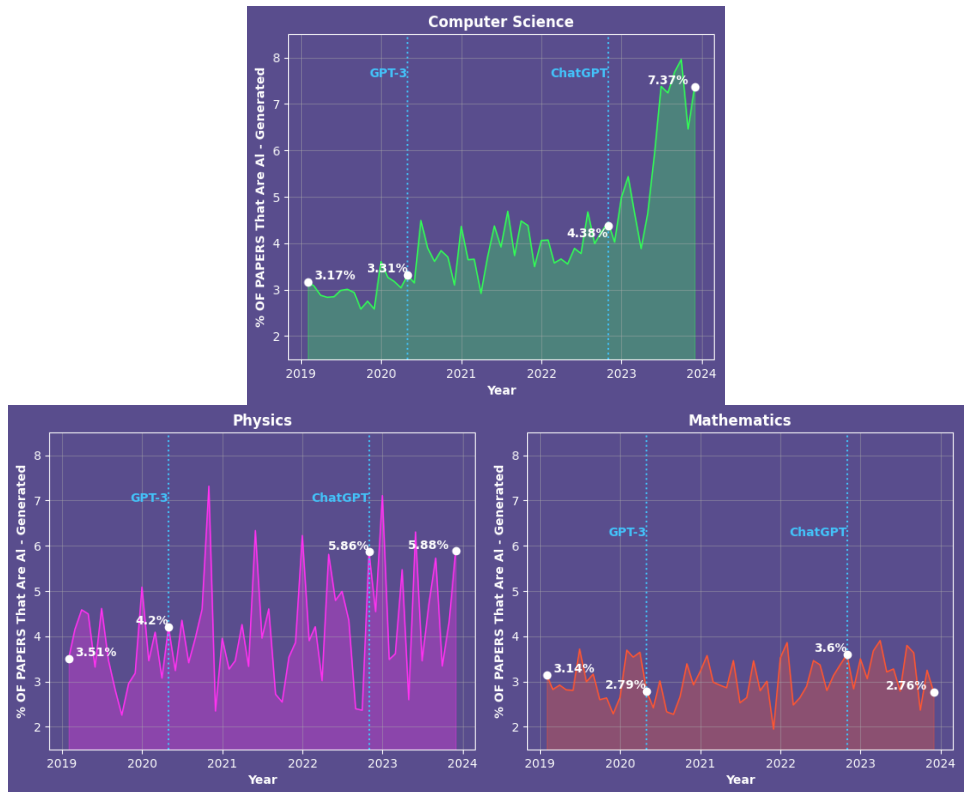

**Figure 6:** Percentage of papers that are ai generated after year 2019 by category with respect to ai score

Our detailed study highlights that AI's role in academic fields, especially in computer science, is quite significant. Starting from the year 2019, there is a gradual increase from 3.17% to 3.31% by the launch of GPT-3. The launch of GPT-3 has a gradual impact, with a slight increase to 4.38% until the launch of ChatGPT. However, after the launch of ChatGPT a significant impact can be observed. The sharp rise reaches up to 7.37% until the end of the year 2023. This trend supports the idea that using AI for writing is impacting various academic fields, with the effect being especially strong in computer science. For Physics and Mathematics, the trend does not provide significant information. The variation in these trends could be influenced by the limitation of AI detection tools, especially in the fields with constant use of numbers and equations.

## 5 Conclusion

The increasing use of AI in research papers is a concern because it might affect the uniqueness and truthfulness of the research. As AI helps write more papers, it's important to think about how to keep the research honest and avoid accidental biases or errors from these systems. Additionally, there's a risk that AI could make research less diverse and limit creative thinking, as it might lead to similar styles and ideas being repeated. This calls for careful monitoring and guidelines to make sure AI helps rather than hinders the quality and diversity of academic work.

**Availability of Data and Materials:**
Data will be provided on request. It is also publicly available.

**Contributions of The Study:**
This study shows how Originlity.AI's AI detection tool can be used to distinguish AI-generated content. It highlights the tool's effectiveness in maintaining content authenticity and points out the tool's role in keeping research original and honest.

**Implications for The Future:**

In the future, being able to find AI-written content will be important for keeping academic work honest and authentic. As AI grows, tools like Originality.AI will be key to making sure research stays authentic and trustworthy.

**References**

[1] S. Baki, R. Verma, A. Mukherjee, and O. Gnawali, "Scaling and effectiveness of email masquerade attacks: Exploiting natural language generation," in *Proceedings of the 2017 ACM on Asia Conference on Computer and Communications Security*, 2017, pp. 469–482.

[2] K. Shu, S. Wang, D. Lee, and H. Liu, "Mining disinformation and fake news: Concepts, methods, and recent advancements," *Disinformation, misinformation, and fake news in social media: Emerging research challenges and opportunities*, pp. 1–19, 2020.

[3] H. Stiff and F. Johansson, "Detecting computer-generated disinformation," *International Journal of Data Science and Analytics*, vol. 13, no. 4, pp. 363–383, 2022.

[4] N. Dehouche, "Plagiarism in the age of massive Generative Pre-trained Transformers (GPT-3)," *Ethics in Science and Environmental Politics*, vol. 21, pp. 17–23, 2021.

[5] A. Radford, J. Wu, R. Child, D. Luan, D. Amodei, and I. Sutskever, "Language models are unsupervised multitask learners," *OpenAI blog*, vol. 1, no. 8, p. 9, 2019.

[6] P. Fortuna and S. Nunes, "A survey on automatic detection of hate speech in text," *ACM Computing Surveys (CSUR)*, vol. 51, no. 4, pp. 1–30, 2018.

[7] "Monthly Submissions." Accessed: Feb. 09, 2024. [Online]. Available: https://arxiv.org/stats/monthly_submissions

[8] Y. Kashnitsky, D. Herrmannova, A. de Waard, G. Tsatsaronis, C. Fennell, and C. Labbé, "Overview of the DAGPap22 shared task on detecting automatically generated scientific papers," in *Third Workshop on Scholarly Document Processing*, 2022.

[9] D. Weber-Wulff et al., "Testing of Detection Tools for AI-Generated Text," *arXiv preprint arXiv:2306.15666*, 2023.

[10] S. G. Burdisso, M. L. Errecalde, and M. Montes y Gómez, "Using Text Classification to Estimate the Depression Level of Reddit Users," *Usando Clasificación de Textos para Estimar el Nivel de Depresión de Usuarios de Reddit*, vol. 21, no. 1, Apr. 2021, Accessed: Feb. 09, 2024. [Online]. Available: http://sedici.unlp.edu.ar/handle/10915/118067

[11] Y. Wu, S. Guan, and G. Wang, "Drug Effect Deep Learner Based on Graphical Convolutional Network," in *Machine Learning and Deep Learning in Computational Toxicology*, H. Hong, Ed., in Computational Methods in Engineering & the Sciences. , Cham: Springer International Publishing, 2023, pp. 83–140. doi: 10.1007/978-3-031-20730-3_4.

[12] A. Akram, J. Rashid, M. A. Jaffar, M. Faheem, and R. ul Amin, "Segmentation and classification of skin lesions using hybrid deep learning method in the Internet of Medical Things," *Skin Research and Technology*, vol. 29, no. 11, p. e13524, 2023, doi: 10.1111/srt.13524.

[13] A. Akram, J. Rashid, A. Jaffar, F. Hajjej, W. Iqbal, and N. Sarwar, "Weber Law Based Approach for Multi-Class Image Forgery Detection," *CMC*, vol. 78, no. 1, pp. 145–166, 2024, doi: 10.32604/cmc.2023.041074.

[14] A. Akram, S. Ramzan, A. Rasool, A. Jaffar, U. Furqan, and W. Javed, "Image splicing detection using discriminative robust local binary pattern and support vector machine," *World Journal of Engineering*, vol. 19, no. 4, pp. 459–466, Jun. 2022, doi: 10.1108/WJE-09-2020-0456.

[15] A. Akram et al., "Enhanced Steganalysis for Color Images Using Curvelet Features and Support Vector Machine," *CMC*, vol. 78, no. 1, pp. 1311–1328, 2024, doi: 10.32604/cmc.2023.040512.

[16] G. Grimaldi and B. Ehrler, "AI et al.: Machines Are About to Change Scientific Publishing Forever," *ACS Energy Lett.*, vol. 8, no. 1, pp. 878–880, Jan. 2023, doi: 10.1021/acsenergylett.2c02828.

[17] H. A. Younis et al., "ChatGPT Evaluation: Can It Replace Grammarly and Quillbot Tools?," *British Journal of Applied Linguistics*, vol. 3, no. 2, Art. no. 2, Oct. 2023, doi: 10.32996/bjal.2023.3.2.4.

[18] C. Chaka, "Detecting AI content in responses generated by ChatGPT, YouChat, and Chatsonic: The case of five AI content detection tools," *Journal of Applied Learning and Teaching*, vol. 6, no. 2, Art. no. 2, Jul. 2023, doi: 10.37074/jalt.2023.6.2.12.

[19] "How Does AI Content Detection Work? – Originality.AI." Accessed: Feb. 09, 2024. [Online]. Available: https://originality.ai

[20] M. Müller, M. Salathé, and P. E. Kummervold, "COVID-Twitter-BERT: A natural language processing model to analyse COVID-19 content on Twitter," *Frontiers in Artificial Intelligence*, vol. 6, 2023, Accessed: Feb. 09, 2024. [Online]. Available: https://www.frontiersin.org/articles/10.3389/frai.2023.1023281

[21] "AI Content Detector Accuracy Review + Open Source Dataset and Research Tool – Originality.AI." Accessed: Feb. 09, 2024. [Online]. Available: https://originality.ai

[22] A. Akram, "An Empirical Study of AI-Generated Text Detection Tools," *Advances in Machine Learning & Artificial Intelligence*, vol. 4, no. 2, pp. 44–55, Oct. 2023.

[23] "arXiv Dataset." Accessed: Feb. 09, 2024. [Online]. Available: https://www.kaggle.com/datasets/Cornell-University/arxiv